# Phase transition between the Bragg Glass and a disordered Phase in $Nb_3Sn$, detected by 3$^{rd}$ harmonics of the AC magnetic susceptibility

Maria G. Adesso[*,†], Davide Uglietti[*], René Flükiger[*]
[*]*Institute of Applied Physics-GAP, University of Geneva,
Quai de l'Ecole de Médecine, 1211 Geneva, Switzerland*

Massimiliano Polichetti[†], Sandro Pace[†]
[†]*Department of Physics, SUPERMAT, INFM, University of Salerno,
Via S. Allende 84081 Baronissi (Salerno) Italy*

(Dated August 3, 2005)

We report additional experimental evidences about the presence of an universal behavior in the Field-Temperature Phase Diagram of Type II Superconductors. This behavior is characterized by a phase transition in the vortex matter between the disordered and the Bragg Glass phase. The experimental detection of a Peak Effect phenomenon has been proved to be strictly connected to the existence of this phase transition. In this paper, we show the first observation of a Peak Effect in the compound $Nb_3Sn$, by using 1$^{st}$ harmonics of the AC magnetic susceptibility. Peak Effect has been detected at fields between 3T and 13T, whereas it is not observable at higher fields. This seems to be in contrast with the theoretical predictions of such a phase transition at all fields and, therefore, with the predicted universality in the magnetic behavior of the Type II superconductors. Nevertheless, by measuring the 3$^{rd}$ harmonics of the AC susceptibility, this phase transition has been detected up to our highest available field (19 T), thus demonstrating the necessity of the higher harmonics analysis in studying these topics and moreover proving the validity of the theoretical predictions.

Magnetic flux penetrates a type II superconductor in form of vortices, which distribute in a regular lattice in absence of defects [1, 2]. Defects affect the lattice ordering, but prevent the dissipations associated to the vortex movement [2]. An universal field-temperature phase diagram has been supposed for all the type II superconductors with point defects. In this phase diagram a transition in the vortex lattice has been predicted, between a disordered phase and the Bragg Glass Phase [3, 4]. This latter is characterized by a quasi long range order and a perfect topological order, in which the vortex lattice stills survives despite the presence of the pinning [3], as it has also been experimentally evidenced [5]. The disordered phase has been supposed to be again a glass phase ("multidomain glass"), but with a topological disorder at the largest length scales [4]. The critical current density, $J_c$ [6], generally decreasing with increasing temperature and/or magnetic field [2], shows a local maximum, known as Peak Effect [7], when the Disordered/Bragg Phase transition occurs. For this reason, the observation of the Peak Effect has been widely used to detect this phase transition [3]. The Peak Effect has been evidenced in various classes of type II superconductors, e.g. low-$T_c$ [8-12], high-$T_c$ [13-15], boro-carbides [16] and $MgB_2$ [17]; however, it has not been observed in $Nb_3Sn$ up to the present day, to the best of our knowledge.

$Nb_3Sn$ crystallizes in the A15 type structure and is actually the most used material in the manufacturing of superconducting magnets at very high fields [18]. In the present work, measurements have been performed on a high quality $Nb_3Sn$ single crystal, furnished by M. Toyota [19] (Dimensions ≈ 3 x 1.32 x 0.43 $mm^3$). The sample was characterized by $T_c$ = 18.2 K, $\Delta T_c$ = 0.1 K, $T_m$ = 38.84 K. A value of $H_{c2}(0)$ = 22.2 T was extrapolated using the GLAG theory [20].

The Peak Effect measurements reported in literature were performed using various experimental techniques [7, 10, 21-25]. The technique used in the present paper is based on the observation of a dip in the real part of the 1$^{st}$ harmonics of the AC magnetic susceptibility, $\chi_1'$ [23]. The $\chi_1'$-dip indicates that the capability of superconductor to exclude the magnetic flux is non monotonic with the temperature. From the Bean critical state model [6], it follows that the dip is directly related to a peak in $J_c$. Measurements of the harmonics of the AC susceptibility ($\chi_{AC}$) have been widely used to study magnetic properties [2, 23,24,26,27], and in particular the vortex dynamics [27] in type II superconductors. If the magnetic response is linear, only the 1$^{st}$ harmonics can be detected, whereas non-linear magnetic properties are associated to the existence of higher harmonics [6,

27]. In this work, we used for the first time the measurements of 3rd harmonics to investigate the phase transition between the disordered and the Bragg Glass phase. This technique allowed to extend the detection of this phase transition to considerably high fields, being only limited by the available field, 19 T.

In our experiments, a home made susceptometer has been used to measure both 1st and 3rd harmonics of $\chi_{AC}$ as a function of temperature, at DC magnetic fields up to 19 T, parallel to the AC field, approximately parallel to the longer dimension of the sample. Measurements have been performed at fixed frequency ($\nu$ = 107 Hz) and amplitude ($h_{AC}$ = 128 Oe) of the AC magnetic field. In Fig. 1, evidence of the Peak Effect is shown, based on the real part of the 1st harmonics, $\chi_1'(T)$. Three different behaviors can be distinguished as a function of the DC magnetic field:

a) 0 T ≤ H < 3 T: the superconducting transition widens with increasing fields, and no dip has been evidenced in the real part of the 1st harmonics.

b) 3 T ≤ H ≤ 13 T: a dip can be observed, which becomes more pronounced up to H = 7 T, its height decreasing for higher fields.

c) H > 13 T: above 13 T the Peak Effect cannot be detected anymore in the 1st harmonics measurements; the superconducting transition is sharper than in a).

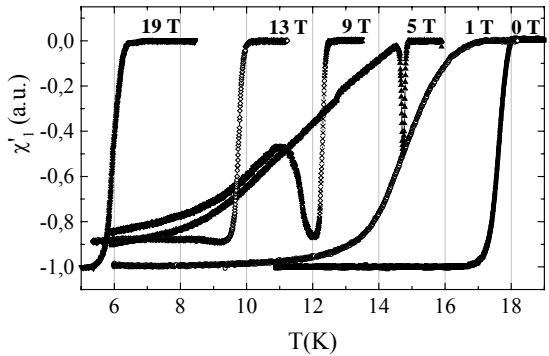

FIG. 1. Detection of the Peak Effect from the real part of the 1st harmonics of the AC magnetic susceptibility, $\chi_1'$, as a function of the temperature, T, at various DC magnetic fields. The Peak Effect has only been observed in the field range between 3T and 13T.

The corresponding H-T phase diagram (shown in Fig. 2) has been obtained by plotting the critical temperature ($T_c$) and the peak temperature ($T_p$), i.e. the temperature corresponding to the local minimum in the dip, at various DC magnetic fields. In the H-T region below the $T_p$ line the vortex lattice still survives (Bragg glass) [3, 5], whereas the disorder destroys the lattice in the temperature range between $T_p$ and $T_c$.

It can be seen that the $T_p$ line stops at a certain point in the high field/low temperature region of the phase diagram. This indicates that the observation of the disordered/Bragg Glass phase transition by means of 1st harmonics measurements is limited to an intermediate field/temperature region. A very similar behavior was earlier reported in the $V_3Si$ system [25], another A15 type compound: the Bragg Glass Phase extends without a transition up to the vortex liquid state, for high magnetic fields (H > 7 T). Nevertheless, the theoretical models [3] predict that this transition should be present, even at higher magnetic fields.

For the sake of clarity, in Fig.2 we did not show the onset dip temperature ($T_{pOn}$) [24, 31], being not associated to a thermodynamic transition [29]. Nevertheless, it remains an open question whether the region between $T_p$ and $T_{pOn}$ is be due either to the introduction of the vortex disorder at the sample edges [30, 32] or to a real coexistence of the two vortex phases, originated by superheating or supercooling phenomena [31].

By magnetic measurements, we have no information about the order of the transition. We detected magnetic history effects (which will be reported elsewhere), for temperature lower than the peak temperature ($T_p$), as in other type II superconductors [24, 25, 28], but they are not a proof of a first order transition. Recent calorimetric measurements [29], in fact, show that the latent heat is zero around $T_p$, suggesting that it cannot be a first order transition.

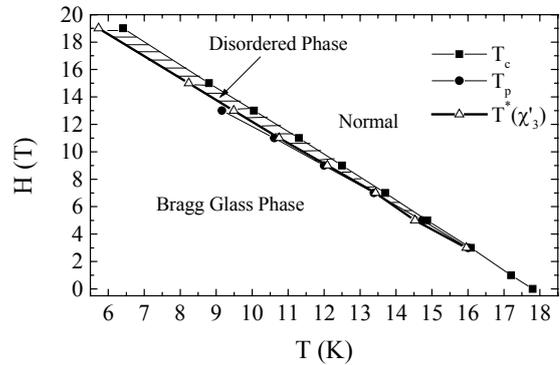

FIG. 2: H-T phase diagram from 1st and 3rd harmonics measurements, obtained by plotting the critical temperature, $T_c$, and the peak temperature, $T_p$, extracted from the $\chi_1(T)$ curves and $T^*(\chi_3')$, associated to the Peak Effect, in the real part of the 3rd harmonics (see Fig. 3). We identified this latter line as the phase transition between the Bragg Glass and the disordered phase.

By means of 3rd harmonics measurements, it was possible to detect the disordered/Bragg Glass phase transition even at higher magnetic fields, this verifying the theoretical predictions of Ref. [3]. The real part of the 3rd harmonics, as a function of temperature,

$\chi_3'(T)$, is shown in Fig. 3 for various magnetic fields. At low fields (H < 3T), where a dip was not observed in $\chi_1'(T)$, the data are in qualitative agreement with those obtained by using the Bean model [6, 27].

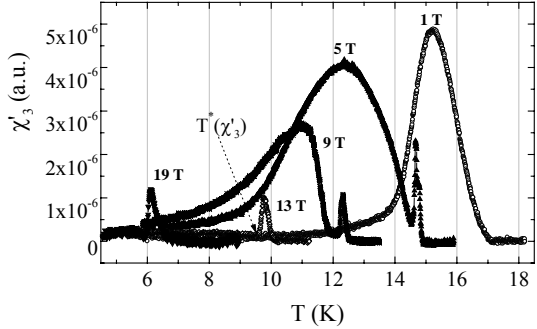

FIG. 3. Real part of the 3$^{rd}$ harmonics as a function of temperature, at various DC magnetic fields.

Indeed, the real part is characterized by positive values while the imaginary part, not reported here, shows a minimum near T$_c$. A new signal around T$_c$ appears in $\chi_3'(T)$ when the Peak Effect is observed. It is almost independent of the DC magnetic field, whereas the contribution to $\chi_3'(T)$ being not associated to the Peak Effect decreases when H is increased. At magnetic fields H > 13T only the new signal can be detected. We used this information to extend the H-T phase diagram, up to 19 T. Fig. 2 also shows the onset temperature $T^*(\chi_3')$ of the new signal, in the real part of the 3$^{rd}$ harmonics. It was found that $T^*(\chi_3')$ corresponds to the peak temperature in the 1$^{st}$ harmonics of the AC magnetic susceptibility. It is remarkable that $T^*(\chi_3')$ can also be individuated in the high field/low temperature region, where the dip was no longer detected in the first harmonics measurements. We verified (see Fig. 4) that $T^*(\chi_3')$ is almost independent of the external parameters (frequency and amplitude of the AC magnetic field, angle α between the sample and the DC magnetic field). We identified the $T^*(\chi_3')$ line in the H-T phase diagram as the transition between a disordered and the Bragg glass phase. In such way, the non-linear signals due to the two different phases can be clearly distinguished in the 3$^{rd}$ harmonics curves. Namely, for T > $T^*(\chi_3')$ there is the peak associated to the disordered phase and for T < $T^*(\chi_3')$ the Bragg Glass phase signal can be detected. Their shapes and dependences on the external parameters can be very different, depending on the vortex dynamics and the pinning properties, due to the fact that the 3$^{rd}$ harmonics is very sensitive to these variations [27]. Detailed theoretical analysis should be necessary to investigate them. Fig. 3 also shows that, at high magnetic fields, the intensity of the 3$^{rd}$ harmonics signal is zero in the temperature range T < $T^*(\chi_3')$, i.e. within the Bragg Glass Phase. The fact that the higher harmonics vanish, implies that the magnetic response in the Bragg Glass phase is linear. A tendency of the Bragg Glass phase towards linearity for increasing DC magnetic fields can be better evidenced by the analysis of the 3$^{rd}$ harmonics Cole-Cole plots, $\chi_3''$ vs $\chi_3'$ (see Fig. 5), i.e. the imaginary part of the 3$^{rd}$ harmonics plotted as a function of the real part [33].

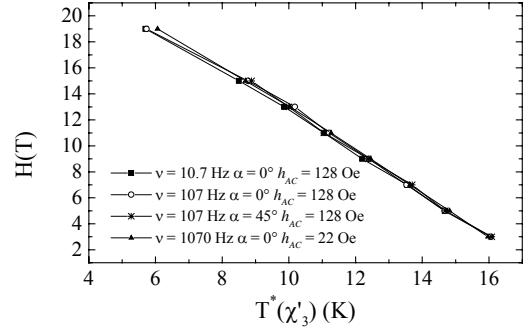

FIG. 4. $T^*(\chi_3')$ measured for various amplitudes, $h_{AC}$, and frequencies, ν, of the AC magnetic field, at the angles α = 0 and 45° between sample and external field. $T^*(\chi_3')$ is almost unaffected by the change of these parameters, thus being a valuable candidate for the transition temperature between the disordered and the Bragg Glass Phase.

At low fields these plots are qualitatively in agreement with those obtained analytically by the static Bean Critical State model (inset of Fig. 5) [6, 33], which predicts for $\chi_3''$ vs $\chi_3'$ a closed loop in the right semi-plane. For increasing fields, the Peak Effect appears, which is reflected by a decrease of the Cole-Cole plot area and a no-loop structure. At higher fields, the Peak Effect in the 1$^{st}$ harmonics is no longer detected and the loop structure in the 3$^{rd}$ harmonics Cole-Cole plots appears again. Its area is almost field independent, and is smaller than at low fields, corresponding to a decrease of the non-linear magnetic response.

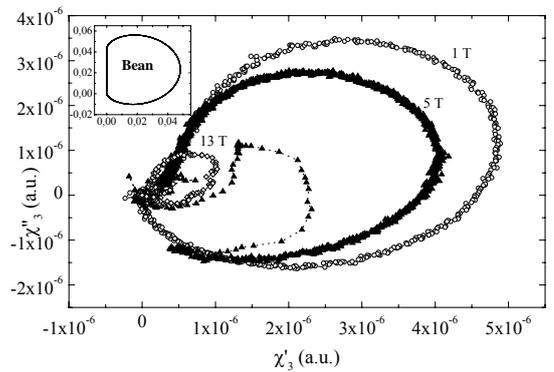

FIG. 5. Cole-Cole plots, defined by the variation of $\chi_3''$, as a function of $\chi_3'$, at various DC magnetic

fields. In the inset, the 3$^{rd}$ harmonics Cole-Cole plots analytically computed by using the Bean model.

In conclusion, a Peak Effect was detected for the first time in the compound Nb$_3$Sn, in a high quality single crystal. Both the Peak Effect phenomenon and the magnetic response in the Bragg Glass phase were studied by means of the 1$^{st}$ and the 3$^{rd}$ harmonics of the AC magnetic susceptibility. It was confirmed that the 3$^{rd}$ harmonics measurements are a more valuable tool for detailed studies of the H-T phase diagram than the 1$^{st}$ harmonics measurements. In particular, they allowed to extend the observation of the Bragg Glass/disordered phase transition to magnetic fields H > 13 T, where the 1$^{st}$ harmonics failed to detect it. The fact that this transition has been observed in Nb$_3$Sn furnishes a previously missing confirmation of the universality of the phase diagram theoretically predicted by Giamarchi et al. [3] in type II superconductors. Finally, our results suggest that 3$^{rd}$ harmonics of $\chi_{AC}$ can be used to individuate this transition in all type II superconductors, possibly also in materials where Peak Effect was not detected so far.

We are grateful to Dr. M. Toyota for sending the Nb$_3$Sn single crystal. We thank Dr. T. Giamarchi for detailed discussions about the Bragg Glass phase. Thanks to Dr. H. Kupfer, Dr. F. de la Cruz, Dr. A. P. Malozemoff, Dr. R. Lortz, Dr. N. Musolino and Dr. G. Deutscher for discussions about Peak Effect and Bragg Glass phase. We are particularly grateful to Dr B. Seeber, R. Moudoux and A. Ferreira for active helping in the laboratory.